\documentclass[twocolumn,showpacs,preprintnumbers,amsmath,amssymb]{revtex4}

\usepackage{graphicx}
\usepackage{dcolumn}
\usepackage{bm}
\begin{document}

\title{Stochasticity of Road Traffic Dynamics:\\
  Comprehensive Linear and  Nonlinear Time Series
  Analysis on High Resolution Freeway Traffic Records}

\author{Helge Siegel}
\altaffiliation{IZKS, University of Bonn,
 Meckenheimer Allee 126, D-53115 Bonn, Germany}
 \email{helgesiegel@yahoo.com}

\author{Denis Belomestnyi}
\altaffiliation{Weierstrass Institute for
Applied Analysis and Stochastics,
Mohrenstr. 39,
D-10117 Berlin
Germany}
\email{belomest@wias-berlin.de}

\date{\today}

\begin{abstract}
 The dynamical properties of road traffic time series
 from North-Rhine Westphalian motorways are investigated.
 The article shows that road traffic dynamics 
 is well described as a persistent stochastic 
 process with two fixed points representing 
 the freeflow (non-congested) and the congested
 state regime. These traffic states have
 different statistical properties, with
 respect to waiting time distribution,
 velocity distribution and autocorrelation.
 Logdifferences of velocity records reveal 
 non-normal, obviously leptocurtic distribution.
 Further, linear and nonlinear phase-plane
 based analysis methods yield no evidence for
 any determinism or deterministic chaos to 
 be involved in traffic dynamics on shorter 
 than diurnal time scales.
 Several Hurst-exponent estimators indicate
 long-range dependence for the free flow state.\\
 Finally, our results are not in accordance to the 
 typical heuristic fingerprints of self-organized 
 criticality. We suggest the more 
 simplistic assumption of a non-critical 
 phase transition between freeflow and congested traffic.
 \end{abstract}


\maketitle

\section{Introduction}
 Traffic flow prediction, particularly in
 close connection with the avoidance of jams, 
 is a challenging, yet hitherto unreached target.\\
 Until now, due to restricted access to records,
 simulation models provided the predominant
 approach to understand traffic dynamics.
 Several approaches have been developed
 which are based on partial differential equations
 (\cite{helbingmodel},\cite{kerner}),
 or cellular automata models as the widespread
 Nagel-Schreckenberg model
  (\cite{schadschneider}).
 A comprehensive overview of results 
 from time series analysis from real 
 traffic records was published by
 \cite{helbing}.
 In earlier research on the database that
 our study relies on, diurnal, weekly and
 annual cycles in traffic density as well as
 velocity was reported in details by
  \cite{chrobok}.
 Autocorrelation and time-headways of
 traffic records are demonstrated to vary
 state-dependently (\cite{knospe0},\cite{neubert}),
 congested traffic re\-vealing a
 more persistant autocorrelation.\\
 Intuitively, traffic dynamics
 conforms rather to a stochastic than
 deterministic(-chaotic) process.
 A rigorous statistical inference however,
 to the best of our knowledge has not yet
 been achieved.\\

 This paper is organized as follows:
 We first introduce the dynamical phase-plane
 reconstruction from traffic records by
 fundamental diagram and  delay-plot,
 to point up that traffic dynamics
 consist of two heterogeneous states.
 The further analysis focuses on
 separated sections of either free
 flow or congested traffic regimes.\\
 We then turn to phase-plane based methods
 such as correlation integrals, and surrogate 
 based local linear predictions to 
 demonstrate that traffic dynamics on below
 diurnal time scales has a predominantly 
 stochastic nature. \\
 Long-range dependence is tested from several 
 measures. To exclude possible effects of 
 nonstationarity, the latter measure is compared 
 with appropriate phase randomized surrogates.
 Nonlinearity will be discussed by application 
 of the surrogate based time-reversibility test.

\section{Data Analysis}
\subsection{Methods}
\subsubsection{Phase-randomized surrogates}
 In time series analysis, phase-randomized
 surrogate (PRS) time series
 (\cite{theiler})
 can be applied as a version of bootstrapping
 to clarify and quantify statements about
 the presence of nonlinear effects.
 PRS series reveal the same linear
 statistical properties as their
 original and can be produced at will.
 Possible nonlinearities, as nonlinear
 determinism beyond the autocorrelation
 of the original time series will not
 be reproduced by their surrogatization,
 or changed by interpretation as a spectral property.\\
 In summary, PRS time series are
 produced by multiplying the Fourier-spectrum
 of the original records with random phases
 and hereafter performing a backtransformation
 (for details see
 \cite{timmer} or
\cite{kantz}).

\subsubsection{Nonlinear methods}
 In this paper we will make use of linear and
 nonlinear phase-plane based measures such
 as correlation dimension and local linear prediction.
 Such methods are usually applied to time series
 with the intention of identifying the presence
 of nonlinear, possibly chaotic dynamics.
 Since it is hardly possible to formally prove
 the absence of any deterministic property,
 we intend to point out this absence by
 comparing (nonlinear) statistics
 for original data vs. their appropriate
 surrogate substitutes.

\subsection{Records}
  Freeway traffic in North-Rhine Westphalia (Germany)
  is continuously monitored at approximately
  1400 road locations by means of built-in loop detectors.
  For every appearance of a vehicle these detectors record:
 \begin{enumerate}
 \item time,
 \item velocity,
 \item type of vehicle,
 \item length of the vehicle.
 \end{enumerate}
  This study is based on two different
  types of static loop-detector recordings:
  \begin{enumerate}
 \item Single car records:\\
  Only a few exceptional time series have been
  recorded with a notebook PC attached to
  the loop detectors,
 \item minute-ag\-gregated data:\\
  These data are obtained from the same
  loop detectors as single- car data.
  However, instead of immediate recording,
  the samplings are exponentially smoothed
  and aggregated in 1-minute intervals.
\end{enumerate}
  Both single-car and minute aggregated records
  are coarse-grained, since all records are
  denoted in $\lbrack$ ''0'' $\ldots $ ''254''
  $\rbrack$, while ''255''
  denotes faulty results. Due to their higher
  resolution, single-car data provide rare,
  but the most detailled (and unspoilt) information,
  particularly for short time scales.\\
  For more details of sampling and processing
  read 
  \cite{knospe0}
  and
  \cite{neubert}.
  Both articles provide a detailled introduction
  into practial aspects of road traffic data.

\section{Results}
\subsection{Time course of traffic dynamics}
\begin{figure}
 \begin{center}
 \includegraphics[height=30mm]{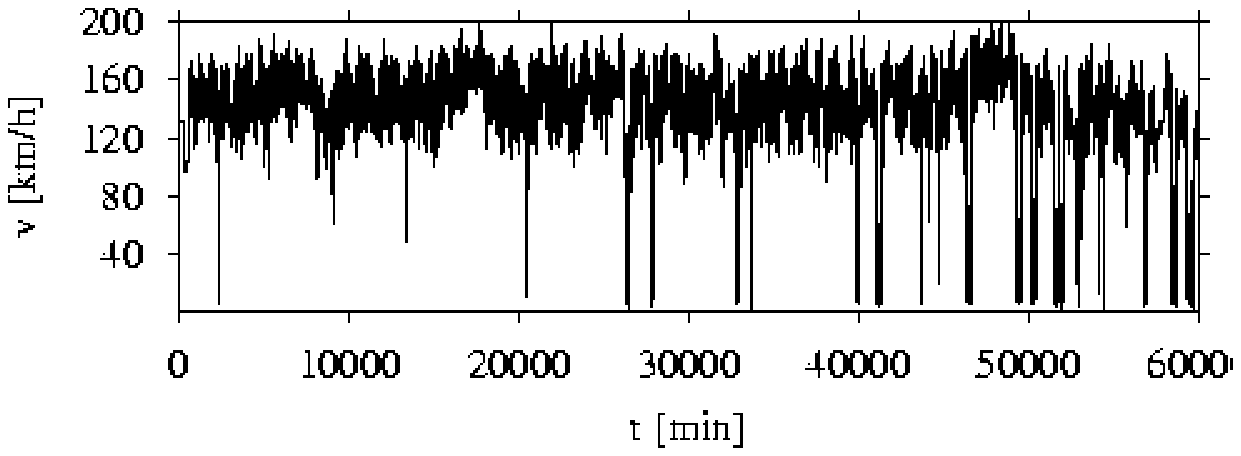}(a)\\
  \end{center}
 \includegraphics[height=30mm]{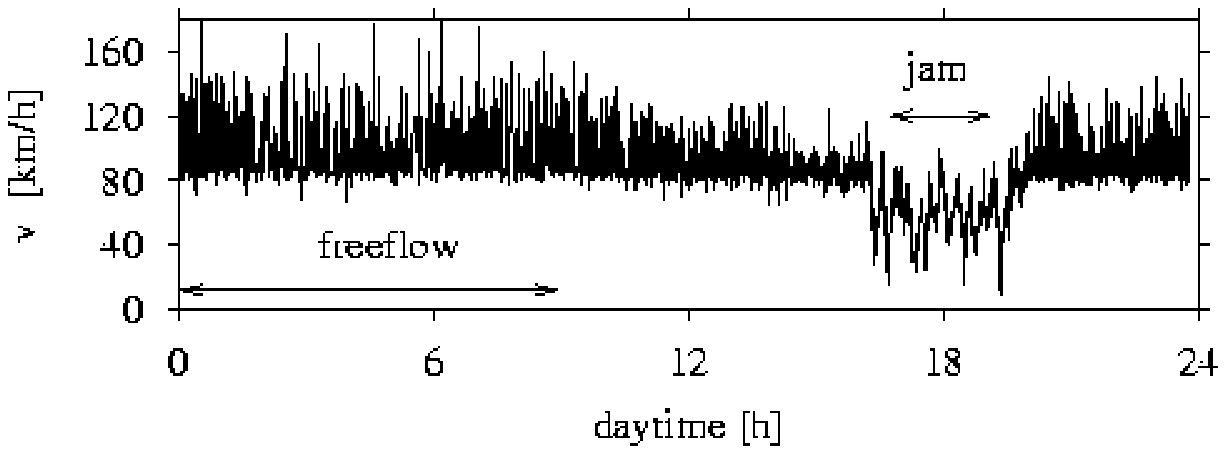}(b)\\
 \caption{\label{siegel_fig01}
 velocities, \newline a) minute aggregated
 records, covering 60.000 min $\approx$ 1000 h
 $\approx$ 40 d.\newline
 (b) 1 day of single car data comprising a jam event,
 arrows indicate sections of jam and freeflow traffic
 state that will be analyzed in the following.}
\end{figure}
 Fig.~\ref{siegel_fig01} presents a section of
 typical single car freeway traffic records.
 In a), the velocity time series appears more
 or less regularly fluctuating, except for
 occasional abrupt drops in velocity,
 (b) is a one-day sequence of single-car
 records comprising a jam episode.

\subsubsection{Fundamental diagram}
\begin{figure}
\includegraphics[height=30mm]{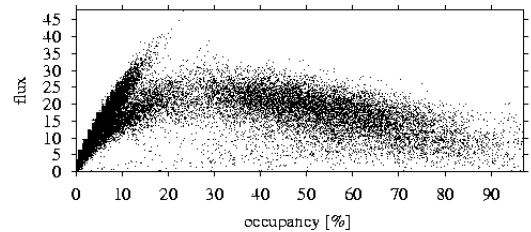}
\caption{\label{siegel_fig03}
 Fundamental diagram: plot of occupancy
 $[$sum of vehicle lengthes per road kilometer$]$ vs.
 flux $[$vehicles per minute$]$
 for minute aggregated traffic records.}
\end{figure}
 In the context of traffic analysis, 
 the fundamental diagram, well known to
 applied scientists, denotes the plot of
 flux vs. occupancy, in most cases graphed
 from smoothed model output data
 (e.g. \cite{wagner}) .\\
 Fig.~\ref{siegel_fig03} was graphed from minute
 aggregated records. Due to the discreteness
 of the latter, in a plot like Fig.~\ref{siegel_fig03},
 some hundreds of thousands of
 data would fall into a few bins.
 To improve the visualization, we added
 uniformly distributed independent random
 noise  $\lbrack -0.5 < \xi < 0.5 \rbrack$
 (the noise level scales below the
 resolution of the signal) to the data.
 Fig.~\ref{siegel_fig03} gives an impression
 of traffic dynamics, that undergoes
 transitions between two attractive regions
 representing freeflow and jammed state.
 Whereas the freeflow regime (high velocity, low
 occupancy) is situated transversally on the left
 hand side, the congested state associates
 with a larger realm of points in the
 center and on the right hand side.

\subsection{Delay plot}
\begin{figure}
 \includegraphics[height=40mm]{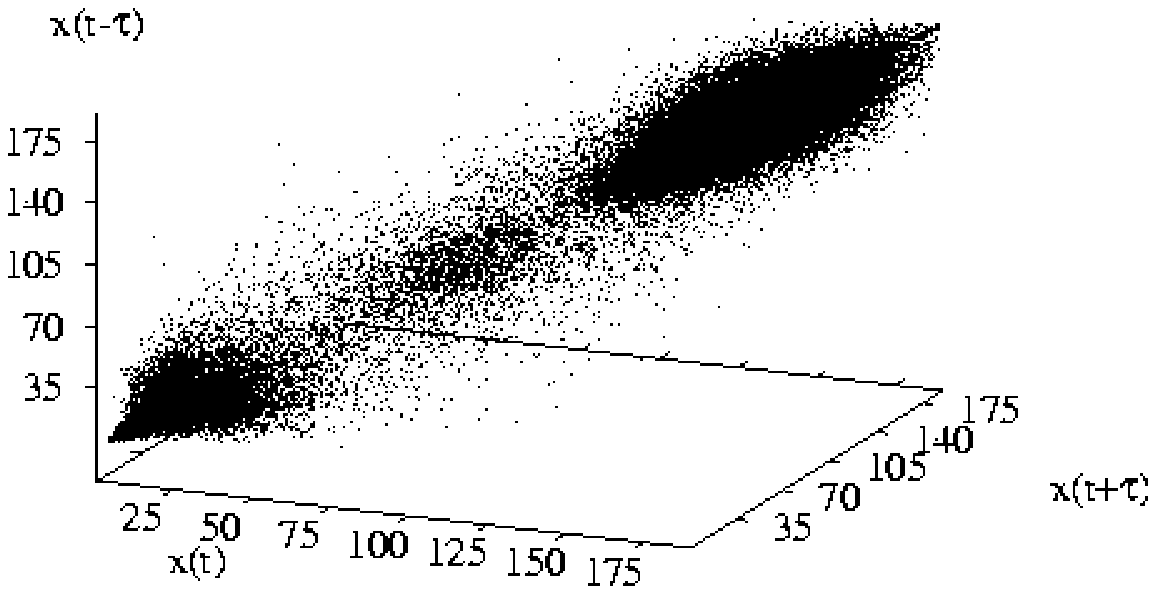}(a)
 \includegraphics[height=40mm]{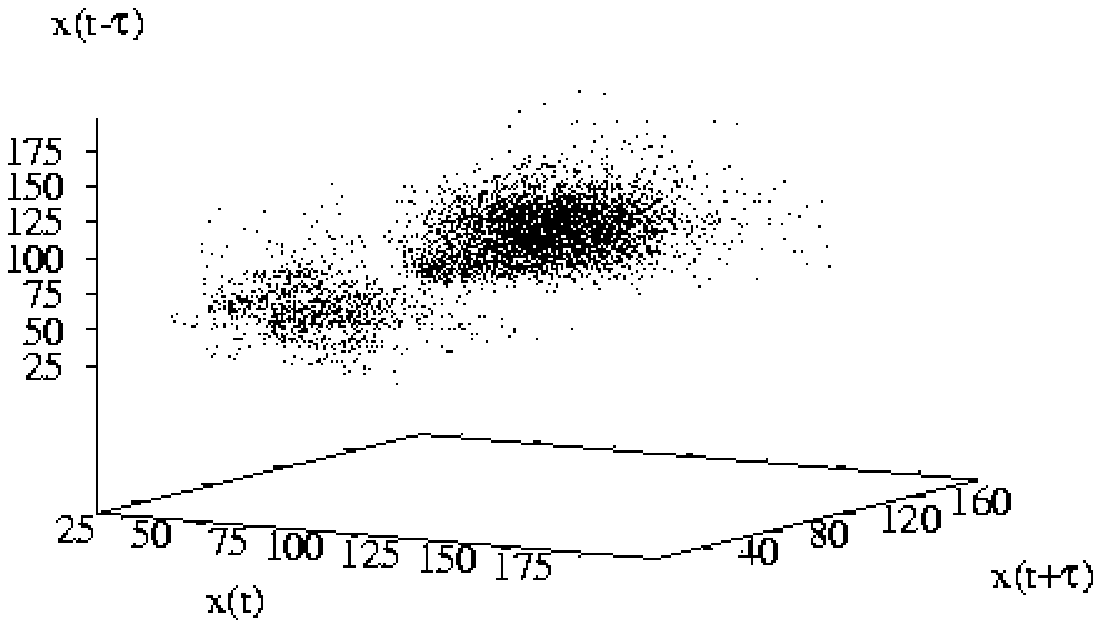}(b)
\caption{\label{siegel_fig04}\newline
           (a) Delay-plot minute aggregated
           velocity data covering one month,
           $\tau = 1$ minute. \newline
           (b) Delay plot of single-car data -
           (1 day), $\tau = 25$ sec.}
 \end{figure}

 Fig.~\ref{siegel_fig03} can be interpreted
 as ''phase plane'' formed from records.
 An alternative, more practicable method
 to obtain a comparable clue on phase plane
 is delay coordinate embedding, which denotes
 a $n$-dimensional plot $x_t$ vs. $x_{t+\tau}$ 
 vs. $\ldots$ vs. $x_{t+n\tau}, n\ll N$
 of a time series $x_t, t = 1, 2,\ldots N$.\\
 The well-known general results by
 Takens \cite{takens} 
 state that  the dynamics of a system recovered
 by delay coordinate embedding are comparable
 to the dynamics of the original system.
 A low dimensional deterministic-chaotic
 attractor thus can be graphed from each of
 its observed variables as a topologically
 equivalent structure to what one would
 obtain from the graph of its variables
 in a sufficiently dimensioned delay plot.
 Since there is no straightforward way to
 determine which dimension is sufficiently
 large, several dimensions need to be examined.
 According to \cite{kantz}
 an optimal delay $\tau$ approximately
 corresponds to the empirical autocorrelation
 function (ACF) at $r(\tau) = 1/e$.
 Fig.~\ref{siegel_fig04}a) shows a delay plot
 of the velocity series $x_t$ used in
 Fig.~\ref{siegel_fig03}, here plotted in
 time-delayed coordinates $x(t-\tau)$,
 $x(t)$ and $x(t+\tau)$, $\tau > 0$
 denoting the delay-time.
 In Fig.~\ref{siegel_fig04} (a) and (b)
 the data scatter around two condensed
 regions, which can be identified as
 congested and freeflow traffic.\\
 We present Fig.~\ref{siegel_fig04}
 to visualize a clearer two fixed-point
 structure than in the ''traditional'' plot
 Fig.~\ref{siegel_fig03}. Moreover, though 
 the single car data base is not sufficient
 to obtain a fundamental diagram,
 Fig.~\ref{siegel_fig04} (b) gives an
 indication of comparable dynamics
 in single car data.

\subsubsection{Local stationarity assumption}
 Naturally, the double fixed point structure,
 visualized in figures~\ref{siegel_fig03}
 and~\ref{siegel_fig04}, gives a strong
 indication against stationarity for the
 overall process of traffic dynamics
 that comprises two traffic states.
 In the following we will therefore constrain the
 analysis to selected sections of either freeflow
 or jammed traffic that are indicated by arrows
 in Fig.~\ref{siegel_fig01}. To apply methods that
 require regularly sampled data we transform
 these sections of single-car data equidistant
 by aggregation and linear interpolation,
 expecting that this procedure does not
 have substantial influence on the results.

\subsection{Distribution of intervals between consecutive events
               (time-headways)}
\begin{figure}
\includegraphics[height=30mm]{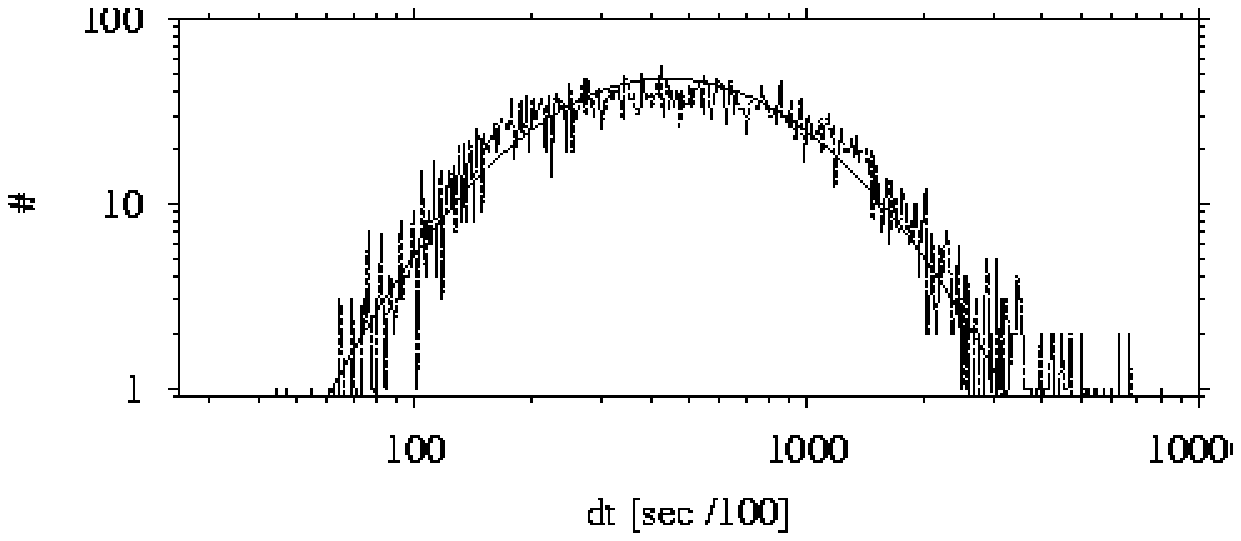}(a)
\includegraphics[height=30mm]{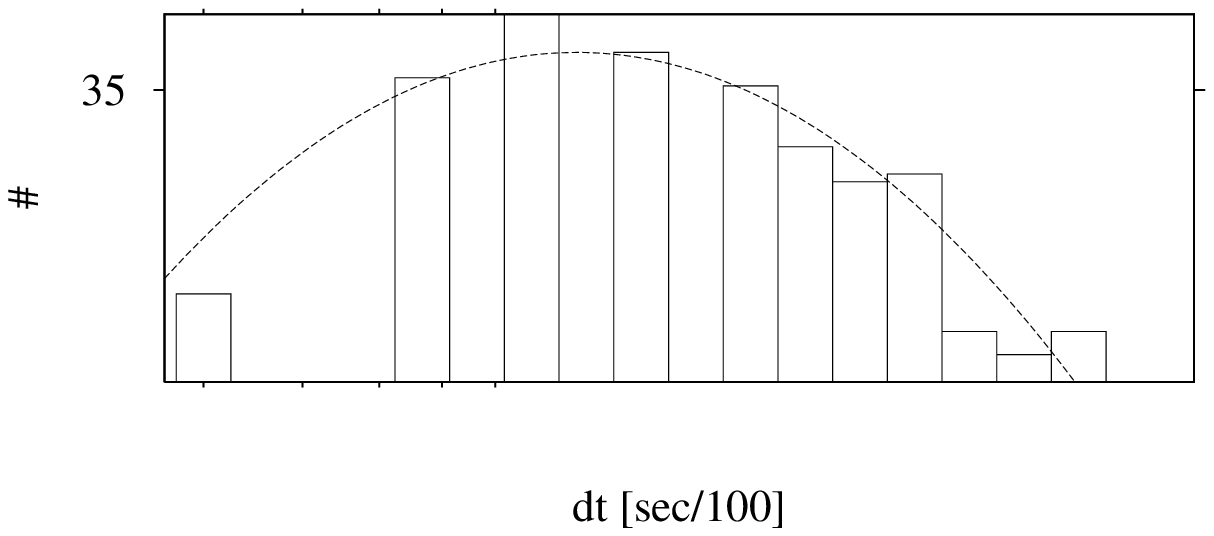}(b)
\caption{\label{siegel_fig06}
        (a) Time interval histogram of
            entire single car data series (dashed line),
            fitted lognormal distribution (solid line)).
        (b) Time interval histogram
            during jammed state,
            fitted lognormal distribution (dashed line).}
\end{figure}

      Time-headway distributions from single car data
      have already been reported in \cite{neubert}
      for different traffic states.
      According to our results they reveal an
      approximately lognormal distribution with different
      parameters in dependence of the traffic state
      (Fig.\ref{siegel_fig06} (a) and (b)).
      For freeflow traffic the Kolmogorov-Smirnov
      test statistics
      \begin{equation}
      \hat D = \frac{\vert x_{min}-x_{max}\vert}{n=500} = 0.0000804\label{eqna}
      \end{equation}
      performs below the tabelled value
      \begin{equation}
           D_{\alpha = 0.001} = \frac{1.949}{\sqrt{n=500}} = 0.087.\label{eqnb}
      \end{equation}
      Thus, this test on distributional adaptation does
      not state the rejection of the null-hypothesis of
      lognormal distribution. In Fig.~\ref{siegel_fig06}(a)
      however, a deviation in the right wing
      (reminding to a fat tail) is observed.
      The finite left tail of the distributions
      probably reflects the necessity to keep
      a security  distance between vehicles.
      Data of jammed traffic (Figure~\ref{siegel_fig06}(b))
      are comparably scarce. Little, if
      anything, can be inferred from them.

\subsection{Self Organized Criticality} 
 Previous authors (\cite{paczuski}) already
 suspected that road traffic has a selfsimilar
 nature in the context of the Self- Organized
 Criticality (SOC) models.
 According to such models, increasing
 traffic load would produce a ''critical''
 situation, that, at its critical point,
 occasionally relaxes catastrophically
 (e.g. as sandslides in the sandpile
 model \cite{bak}).
 Close to the critical point, such a system
 generates power law behaviour, observable
 in leptocurtic distributions, slowly converging
 variance, lack of characteristical scales
 and 1/f noise.

\subsubsection{Distribution of velocities and velocity differences}
\begin{figure}
 \includegraphics[height=30mm]{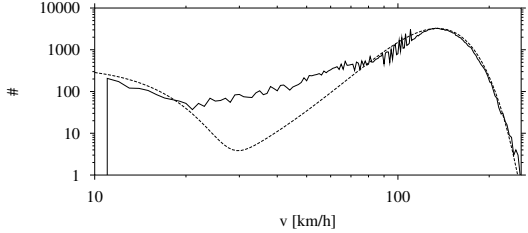}
  \caption{\label{siegel_fig15}
            Histogram of single car velocities of 12 different
            single-car highway traffic data sources (solid line),
            \newline interpretation as addition
            of 2 Gaussian distribution curves
	     (dashed line).}
  \end{figure}

  \begin{figure}
   \includegraphics[height=30mm]{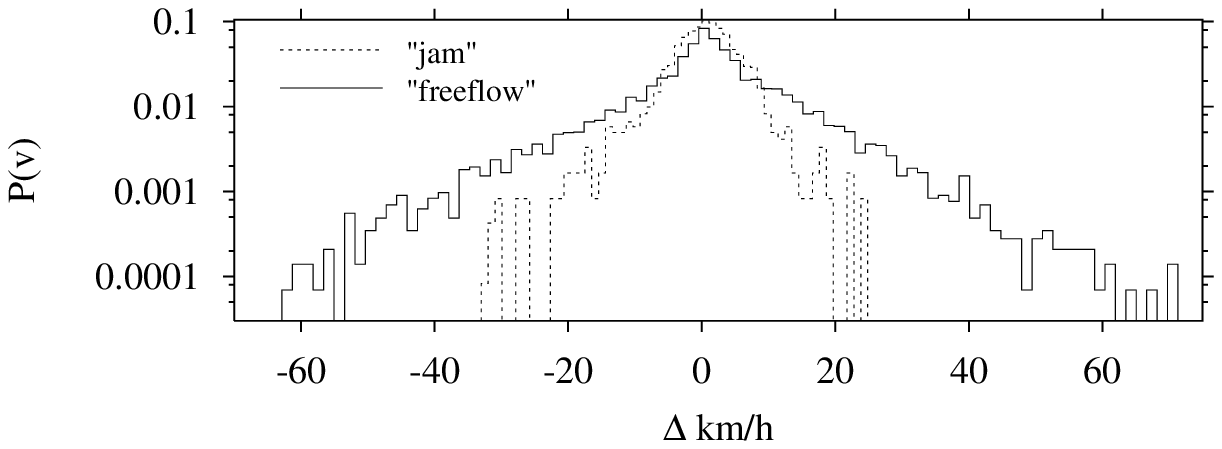}(a)
   \includegraphics[height=30mm]{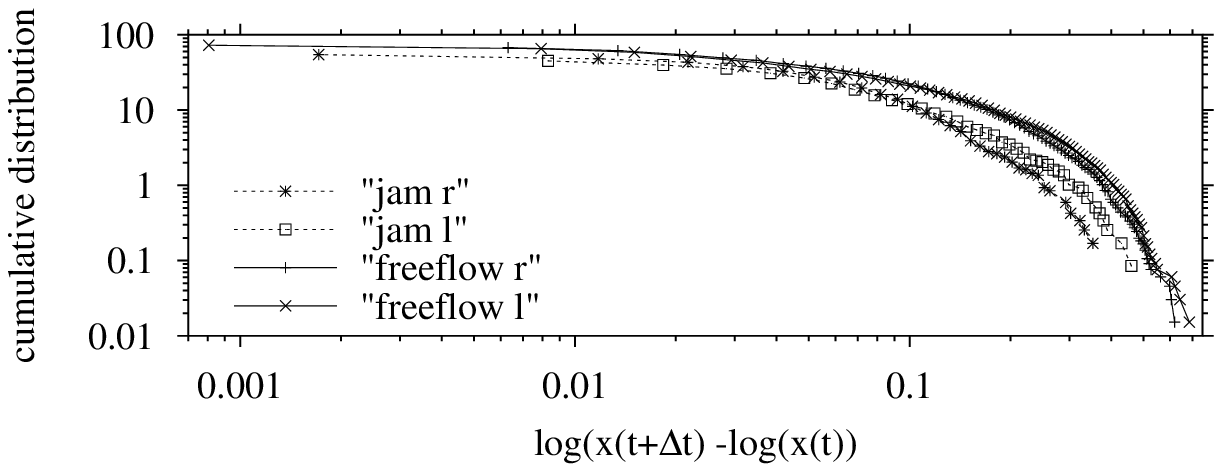}(b)
    \caption{\label{siegel_fig08}
              (a) Histogram of differenced single car velocities
              selected for jam and freeflow epsidodes,
              \newline
              (b) cumulated distribution of logdifferenced
               single car velocities selected for
               left(l) and right (r) wings.}
  \end{figure}

 Fig.~\ref{siegel_fig15} shows the histogram
 of traffic velocities of all single-car data
 (12 different locations). In some locations
 slow congested traffic and jams appear
 as a smaller second peak that, e.g.
 for smaller data quantity, could
 be misinterpreted as fat tail in the
 low-speed end of the empirical velocity
 probability distribution function.\\
 Comparable to the well-known heavy-tailed
 distributions of logdifferenced financial time series,
 in fig.~\ref{siegel_fig08} (a) we observe a clearly
 non-normal distribution in differenced velocity records,
 separated for either jam and free-flow records.
 This holds for logdifferenced data as well (not
 shown here). The histogram looks more
 leptocurtic for free-flow than for jammed
 traffic records. \\
 The plot of the cumulated distribution
 function in double-log coordinates
 provides a clue if the asymptotic behaviour
 of the functional form of the cumulative
 distribution is "visually"
 consistent with a power law,
 \begin{equation}
   P(G > x_*) \approx \frac{1}{x ^{\alpha}} \label{cumul}
 \end{equation}
  where $\alpha $ is the exponent caracterizing the power law decay,
  \begin{equation}
    G(t, \Delta t) \equiv \ln (x_{t+\Delta t})-\ln (x_t) \label{logdiff}
  \end{equation}
 (\cite{plerou}).
 Together with other indications, such 
 a power law can be regarded as a feature 
 which is characteristic for SOC                
 processes (\cite{jensen}).\\
 Conversely, the cumulated distribution of
 differenced velocities,  separated for right
 and left wings of either jam and freeflow
 records, displays no clear scaling region.
 Non-normal distribution as well as lack
 of scaling is also observable for  the larger
 database of minute aggregated records (not
 shown here).

 \subsection{Long-range dependence}
 Scientists in diverse fields observed
 empirically that positive correlations
 between observations which are far apart
 in time decay much slower than would be
 expected from classical stochastic models.
 In time series such correlations are
 characterized by the Hurst exponent $H$.
 They are often referred to as Hurst
 effect or long-range dependence (LRD).
 $0.5<H<1$ reflects long-range positive
 correlations between sequential data.
 $H=0.5$ corresponds to sequential
 uncorrelatedness (known for white noise).
 Brownian motion, the trail of white noise,
 is characterized by $H=1$.\\
 Since long-range dependence (LRD) is defined
 by the autocorrelation function (ACF),
 theoretically, the shape of the ACF provides  
 an indication for LRD in road traffic.
 For LRD series, the ACF at large lags
 should have a hyperbolical shape:
 \begin{equation}
 r(\tau)\propto\tau^{2H-2},\tau\rightarrow\infty \label{eqne}
 \end{equation}
 (\cite{taqqu}).\\
 The practical ability to assure an
 algebraic decay of the ACF however is low,
 making such an approach inviable for
 data analysis. For comparable reasons, from the
 tail of the distribution, additional information
 is hardly obtainable; statistics here are
 generally poor (\cite{carreras}).
 The discreteness of car traffic data additionally
 diminishes the quality of such estimations.

\subsubsection{Hurst-exponent estimation}
 The estimation of the Hurst-exponent
 ($H$) from empirical data is not a simple task.
 Several studies (e.g. \cite{taqqu},\cite{molnar},\cite{laughlin})
 estimate the Hurst exponent $H$ from different measures.\\
 Synthetically generated fractional Brownian
 motion or fractional ARIMA (autoregressive
 integrated moving average) series are
 characterized by a generalized (or global)
 $H$.  Such so called monofractional series are
 known to reveal fluctuations on all time scales.\\
 They will produce unambiguous evidence
 for fractionality, whereas a more
 general class of heterogeneous signals
 exist that are made up of many interwoven
 sets with different local Hurst exponents,
 called multifractional (\cite{latka}).
 It is a frequent experience, that 
 graphical methods to test for LRD 
 show no clear scaling for such series.
 Our own experience is, that weighted sums
 of synthetically generated random walks
 with different characteristical scales
 may as well reveal straight fractional
 scaling in some plots, as crossover behaviour
 according to other methods. Furthermore,
 some methods of $H$-estimation sensitively
 depend on the distribution of the data.\\
 The main criticism against $H$-estimates is
 based on the experience that instationary
 data may, at least in some cases, produce
 estimates that erroneously indicate fractionality.
Thus, we are interested in the robustness of
 $H$-estimators, if possible effects
 of instationarity are excluded.
 Phase-randomized surrogates (PRS)
 based on original traffic records are
 random sequences with the same first
 and second order properties (the mean,
 the variance and the auto-covariance
data, but which are otherwise random.\\
 Since fractionality is a spectral property,
 and PRS fully recover the latter,
 $H$-estimation of PRS hence provides an
 approach to exclude possibly misleading
 effecs of instationarity, albeit not
 to differentiate between monofractional
 and heterogeneous signals.

\begin{figure}
\includegraphics[height=30mm]{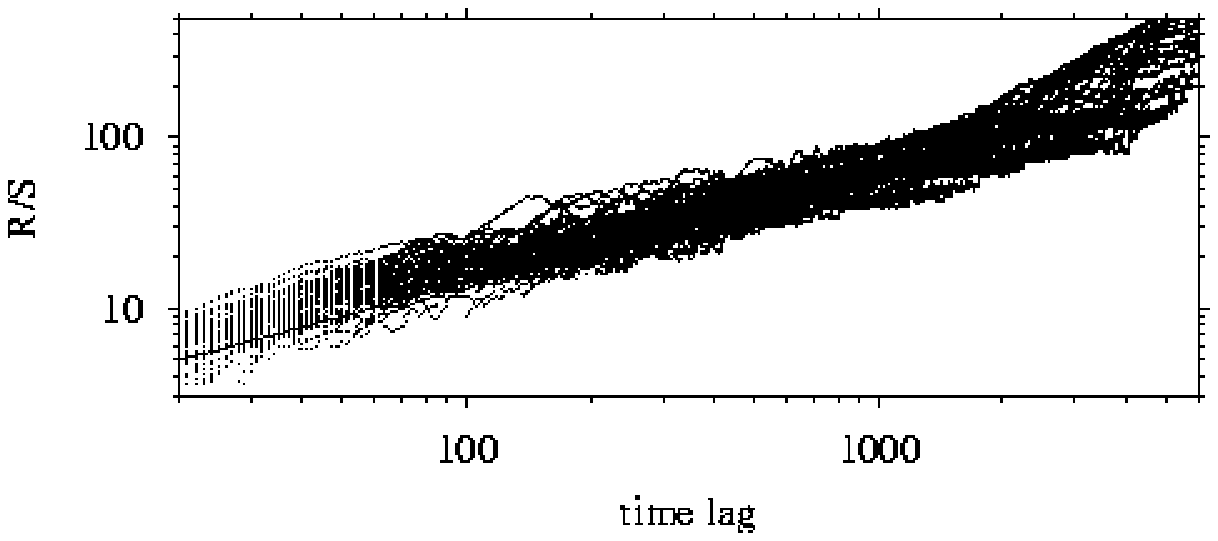}(a)
\includegraphics[height=30mm]{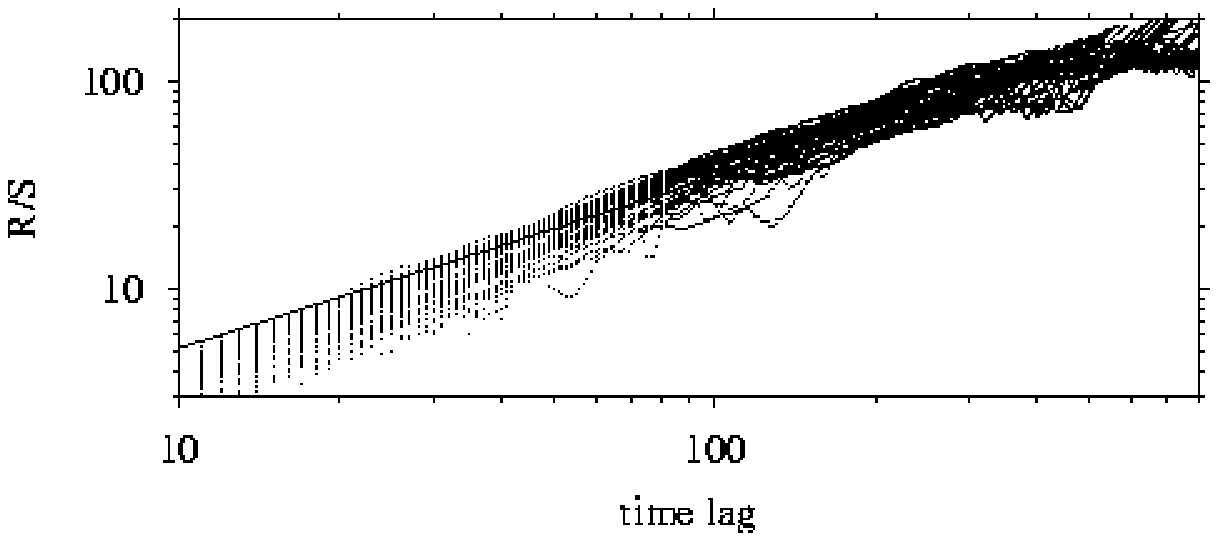}(b)
\caption{\label{siegel_fig11}
           \newline
         R/S pox plot of (a) freeflow and
         \newline (b) congested velocity records,
         approximation of scaling
         by Hurst exponents (a) $H=0.71$,
         (b) $H= 0.81$  (solid lines).}
\end{figure}
\begin{figure}
 \includegraphics[height=30mm]{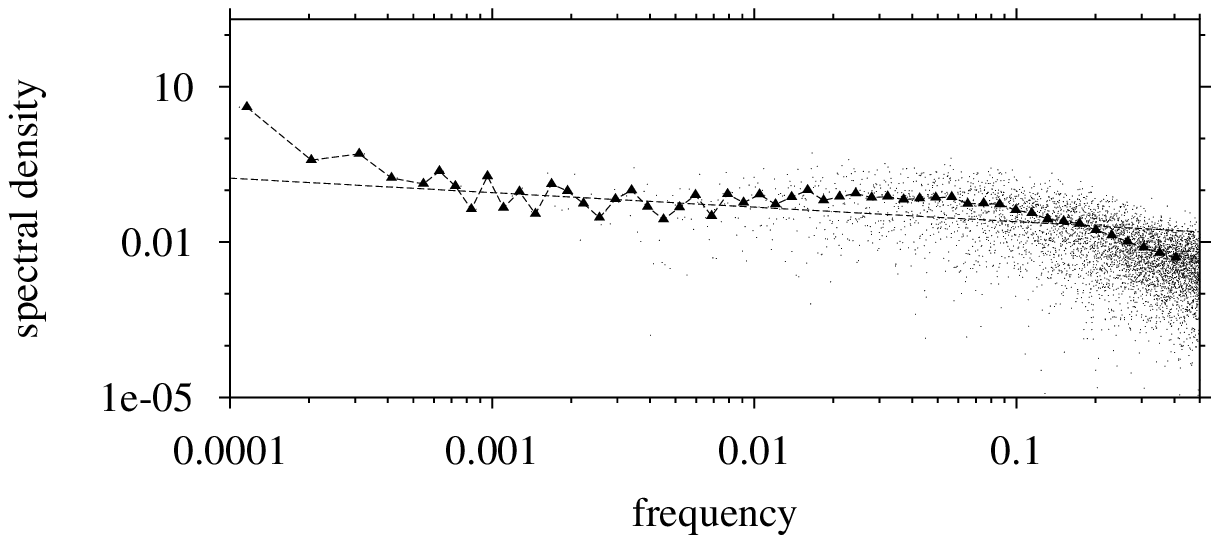}(a)
 \includegraphics[height=30mm]{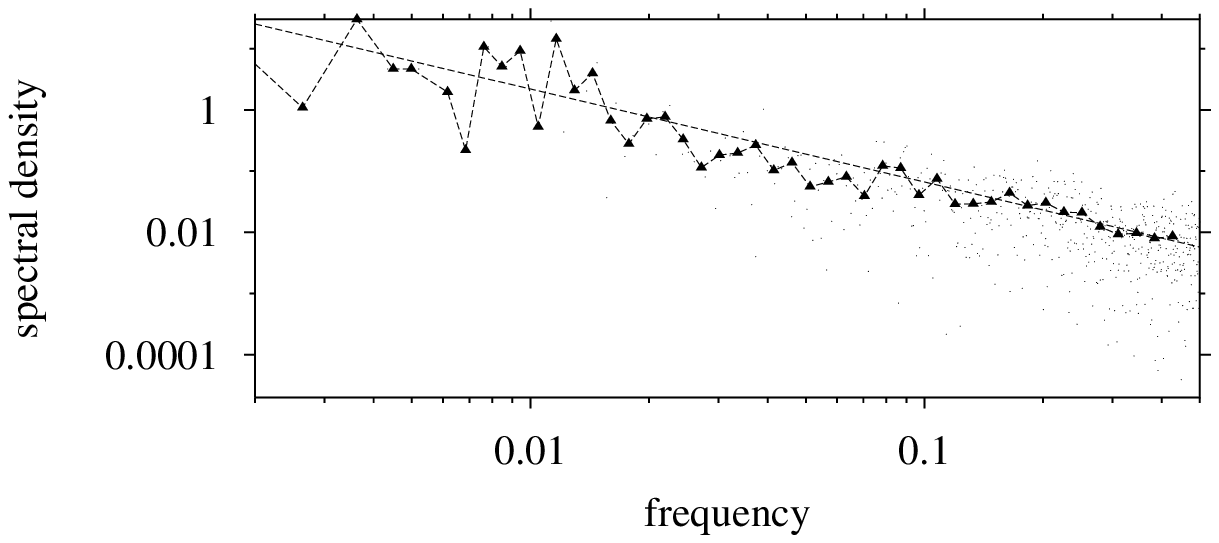}(b)
 \caption{\label{siegel_fig13}
 \newline
 (a) Spectrum from selected freeflow series 
 (dots), exponentially weighted moving average
 (solid line), scale exponents $\alpha
 \approx 0.71,$ (dashed line). \newline
 (b) Spectrum from selected jam series (dots),
 exponentially weighted moving average (solid line),
 scale exponent $\alpha=0.88$ (dashed line)}
\end{figure}
 To obtain reliable H-estimates despite possible
 effects of nonstationarity, we apply a variety
 of the most familiar methods to jam- and freeflow
 traffic records. For a detailled discussion about
 the application of methods to conclude on LRD,
 e.g. aggregated variance method, R/S plot,
 periodogram method and wavelet-based
 Whittle-estimator on nonstationary data
 read \cite{molnar}.
 Detrended fluctuation analysis (DFA)
 (\cite{peng}) denotes the
 root mean square fluctuations
 \begin{equation}
 F(n) = \sqrt{N^{-1}\sum_{k=1}^{N}
 \lbrack y(k)-y_n(k)\rbrack^2}\label{eqnf}
 \end{equation}
 around least squares line trend fits $y_n(k)$
 for equal box length $n$ of the integrated series
 \begin{equation}y(k) = \sum_{i=1}^k (x_i-\bar x),
 \qquad \bar x = N^{-1} \sum^N_{i = 1}.\label{eqng}
 \end{equation}
 A straight line in the double logaritmic plot
 that indicates scaling has the slope $2 H$.
 The method should provide robust estimates
 even for nonstationary time series.
 Table~\ref{fraktab} displays the results. \\
 We also applied the  wavelet-based
 Whittle estimator (\cite{abry}).
 Despite its postulated robustness against
 instationarity, and despite $H$-estimates
 that compare to table~\ref{fraktab},
 we do not show the graphs here,
 since, particularly for the jam series,
 the wavelet-spectrum offers to many
 possibilities of parametrization, as.
 e.g. the choice of the wavelet function,
 octaves etc..\\
 \begin{table}
 \begin{tabular}{||l|l|l|l|l||}
 \hline
\hline
est.              & freeflow   & $\sigma$           & jam            & $\sigma$    \\
 \hline
R/S               & $0.657$    &                    & $0.823$        &            \\
'' for PRS        & $0.656$    & $0.012$            & $0.842$        & $0.016$    \\
a.v.              & $0.761$    &                    & $0.89 $        &            \\
'' for PRS        & $0.779$    & $0.022$            & $0.896$        & $0.041$    \\
a.a.              & $0.522$    &                    & $0.669$        &            \\
'' for PRS        & $0.511$    & $0.002$            & $0.586$        & $0.035$    \\
spc.              & $0.611$    &                    & $0.875$        &            \\
DFA               & $0.685$    &                    & $1.107$        &            \\
'' for PRS        & $0.669$    & $0.005$            & $1.232$        & $0.017$    \\
 \hline
 \hline
 \end{tabular}
 \caption{\label{fraktab}
 Hurst-exponent estimators from traffic
 records for different methods:
 R/S : rescaled range analysis,    \newline
 a.v.: aggregated variance method, \newline
 a.a.: aggregated absolutes, \newline
 spc.: graphical estimation from the spectrum, \newline
 DFA: detrended fluctuation analysis.
 PRS denotes the application of the above methode
 to phase-randomized surrogates, \newline
 $\sigma $ denotes the standard deviation. }
 \end{table}
Fig.~\ref{siegel_fig11} presents the R/S pox plots of freeflow
and jam records. Particularly for freeflow data
exact straight scaling is not observable.
The same, even more, holds for Fig.~\ref{siegel_fig13} a).
In anaogy to the the modified periodogram method
 outlined in \cite{taqqu}, the logaritmically spaced
 spectrum was divided into 60 boxes of equal length.
The least squares ft was performed to averages of  the
data inside these boxes, to compensate for the fact
that most of the frequencies in  the spectrum fall on
the far right, whereas for LRD-investigation the
low frequencies are of interest.
Fig.~\ref{siegel_fig13}(a) gives the strongest 
indication that traffic dynamics can not be 
characterized as monofractional as most of 
the common Hurst-estimators would indicate.

\subsection{Time reversibility test}
\begin{figure}
\includegraphics[height=30mm]{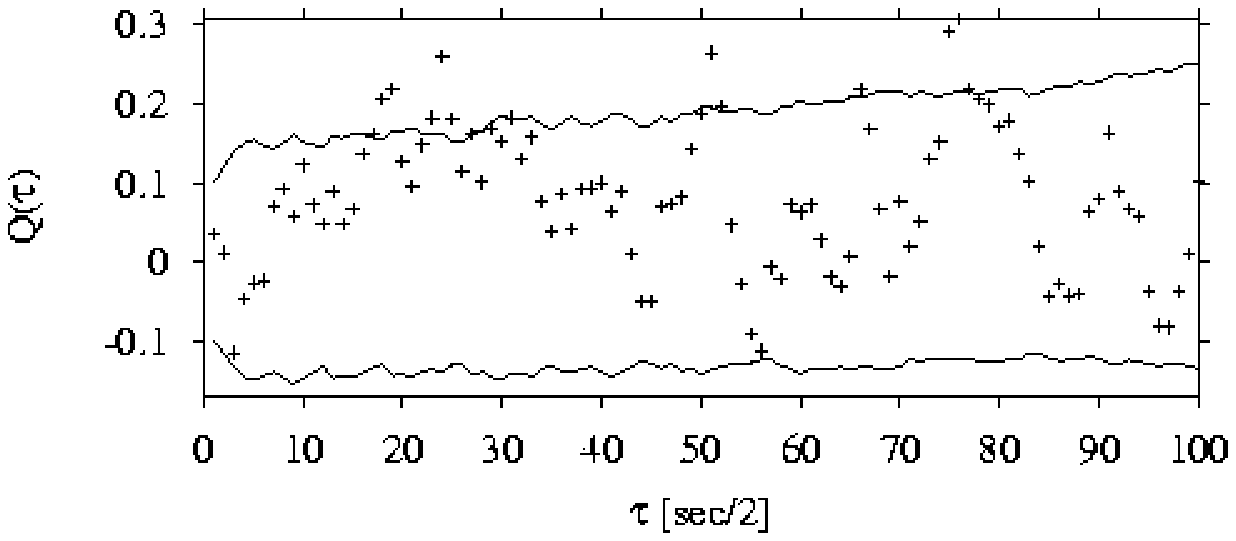}(a)
\includegraphics[height=30mm]{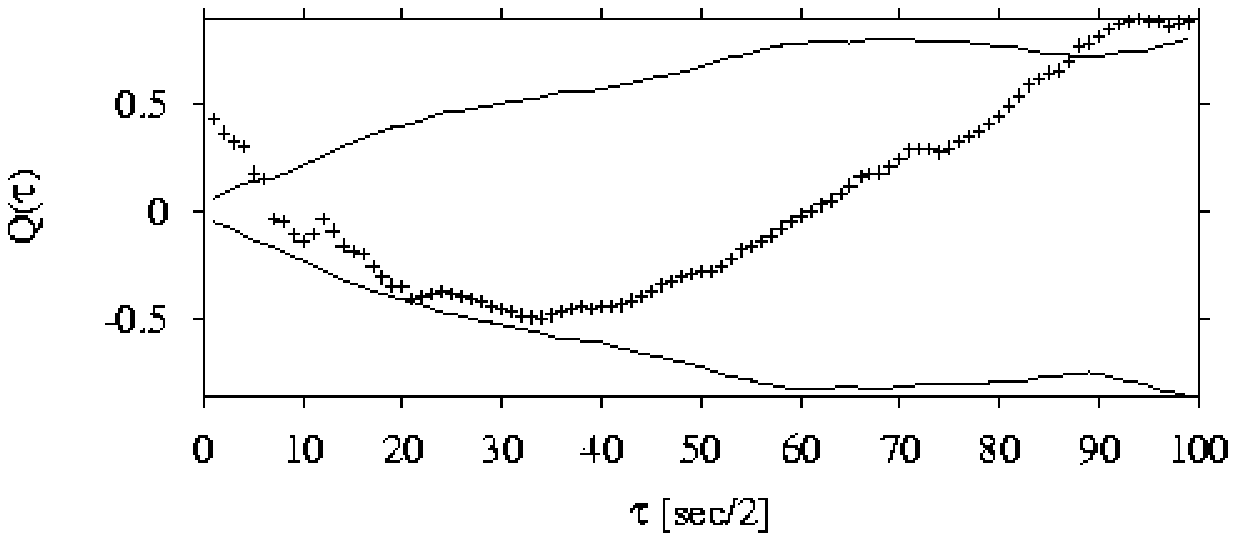}(b)
\caption{~\label{siegel_fig16}
        (a) $Q(\tau)$ from freeflow series (crosses),
           lines indicate confidence limits
           from 500 surrogate realizations.
        (b) $Q(\tau)$ from jam traffic series (crosses),
           lines indicate confidence limits
           from 50 surrogate realizations.}
\end{figure}

 An important property to differentiate
 between linear and nonlinear stochastic
 processes is time-reversibility, i.e.
 the statistical properties are the
 same forward and backward in time.
 From this test one can not judge whether
 the data correspond to any ARMA-model,
 since theoretically, time asymmetry
 might be caused by non-Gaussian
 innovations. Apart from on-ramps, traffic
 dynamics on short time scales anyway
 is unlikely to be substantially
 influenced by external noise.\\
 The following expression is outlined as
 a measure to conclude on time reversibility
 of time series (\cite{theiler}):
 \begin{equation}
 Q(\tau) = \frac{ E \lbrack (x_{t+\tau}-x_t)^3\rbrack
 }{E\lbrack (x_{t+\tau}-x_t)^2\rbrack }\label{eqni},
 \end{equation}
 wherein $\tau$ denotes the delay time
 and $E$ represents the time average.
 The basic idea behind it is to compare
 the time reversibility test statistics 
 of the original data $Q(\tau)$ with 
 confidence bounds from corresponding 
 test statistics $Q_{sur}(\tau)$,
 generated from surrogate series:
 \begin{equation}
 Q_{surr}(c'_\alpha, \tau) < Q(\tau)< Q_{surr}(c_\alpha, \tau) \label{eqnj};
 \end{equation}
 for some critical $c'_\alpha;c_\alpha$. \\
 The results for a surrogate-based test
 are usually reported as significances:
 \begin{equation}
 S(\tau)=\frac{\sqrt{(Q(\tau)-\langle
 Q(\tau)\rangle_{surr})^2}}{\sigma(Q(\tau))_{surr}}\label{eqnk},
 \end{equation}
  where: \\
  \begin{tabular}{ll}
  $ Q(\tau)$               &test statistics,\\
  $ \langle Q(\tau) \rangle_{surr}$      &mean, \\
  $\sigma(Q(\tau))_{surr}$ &standard deviation.\\
 \end{tabular}\\
 The test is based on the assumption that
 the surrogate test statistics for a given
 lag are approximately Gaussian distributed.\\
 The statistical properties of the examined
 surrogate time series are the same forward
 and backward in time. Thus they comply with
 the null-hypothesis of time-reversibility
 which will be tried to reject by the test.\\
 The evaluation of significances for more than
 one lag leads to the statistical problem of
 multiple testing. This has severe implications
 on the probability to reject the null-hypothesis.
 The Bonferroni-correction of the significance
 level must be taken into account:
\begin{equation}
 \hat \alpha= 1 - (1-\alpha)^n \label{multest}.
\end{equation}
wherein $n$ denotes the number of independent tests.
Practically, Bonferroni- corrected confidence bands
render little diagnostic power to detect a violation
of the null-hypothesis. A corrected
significance level, for example,
$1-\hat \alpha = 0.95$ for $100$ independent
tests requires $1-\alpha \approx 0.9995$.
In most cases, however, $Q(\tau)$ is autocorrelated
to an unknown extent, what diminishes the number of
independent tests and, for rejection of the
null-hypothesis, results in a conservative test design.\\
In Fig.~\ref{siegel_fig16}, surrogate-based
time-reversibility tests for jam and
freeflow traffic states are graphed.
Under the assumption of 100 independent
tests for $\alpha=0.001$,
the corrected significance level is:
$\hat \alpha = 1-(0.9999)^{n=100}\approx 0.905,$
which, though not acceptable as safe statistical 
inference,  gives a vague information that freeflow
traffic dynamics is more likely time-irreversible
than time-reversible.
For the jammed state the observed 20 deviations
of the confidence bands gives a comfortably safe
rejection of the null-hyothesis, particularly
for short time scales, but also for larger $\tau$.
Since, even for the naked eye, the test statistics
is substantially correlated, the test provides a safe rejection.
 For the freeflow state, 14 deviations from $H_0$
are also statistically indicative, albeit less
correlation among the test statistics is observable.
Both traffic states thus are likely to reveal
time-irreversible statistical properties.

\subsection{Recurrence plot}
\begin{figure}
\includegraphics[height=40mm]{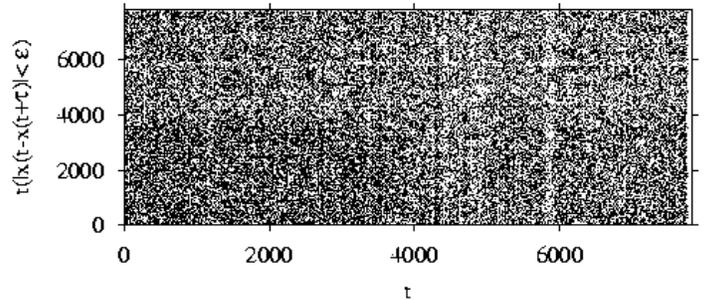}
\caption{\label{siegel_fig18} Recurrence plot of freeflow
           data, $\tau = 1.7 $ seconds,
             $\epsilon = 0.01$, $m = 3.$}
\end{figure}

 For a time series $x_t$ the recurrence plot is a
 two-dimensional graph that is formed from embedded vectors
\begin{equation}
\vec y(t) = (x_t, x_{t+\tau}, \ldots, x_{t+(E-1)\tau}) \label{eqnl}
\end{equation}
 for embedding dimension $E$ and lag $\tau$.
 These vectors are compared if they
 are in $\epsilon$-proximity of another
 $\vec y(t + \Delta t)$.\\
 If
\begin{equation}
\vert\vert\vec y(t)-\vec y(t+\Delta t)\vert\vert
 < \epsilon, \label{recur}
 \end{equation}
 a black point is drawn at $(t, \Delta t)$.
 For each $\epsilon, \tau, m$ (with: $m$ the
 embedding dimension, $\tau$ the time
 lag, $\epsilon$ the variable error distance) an
 individual recurrence plot is obtainable.\\
 Since the differences
 \begin{equation}
 \vec y(t_i)-\vec y(t_j) = c_{i,j} = c_{j,i}
  =\vec y(t_j)-\vec y(t_i)\label{eqnm}
  \end{equation}
 are identical, the plot consists of two symmetric
 triangular graphs along a black
 (since $i=j$) diagonal line.\\
 Except for horizontal and vertical stripes
 (that might reflect temporal (auto-) correlations),
 the recurrence plot of freeflow traffic
 Fig.~\ref{siegel_fig18} is very much 
 in remedy of what one would observe 
 for a recurrence plot of a white noise series.

\subsection{Correlation integral}
 There exist $$ N_T = \frac{1}{2}N(N-1)$$
 independent radii $c_{n,m}$ (since $c_{n,m}=c_{m,n}).$
 The density of a recurrence plot
 as functional of $\epsilon$
 \begin{equation}C(\epsilon, D, \tau)
  = 1/N_T \sum_{t = 1}^{t=N}\sum_{\tau=1}^{n-1}
 \theta(\epsilon -\vert x_t - x_{t-\tau} \vert_E) \label{eqnn}
 \end{equation}
 ($\theta$ denoting the Heavyside step
 function, \\
  with  $\theta (z) = 1,$  for $z > 0, \, \,\, \,
 \theta (z) = 0$ for $z \le 0$) \\
 is called the correlation integral.
 The resulting $C_r(\epsilon)$ is
 sketched in a double logarithmical
 Grassberger-Procaccia plot (\cite{grassberger})
 in dependence of $\epsilon$.\\
 The correlation integral is plotted for varying
 dimension as well as varying $\epsilon$.
 If a noise-contaminated deterministic
 process is regarded, from a sufficient
 embedding dimension, parallel slopes for varying
 dimensions indicate power law scaling in a
 region which is situated above a certain
 $\epsilon$ that represents the noise range.

\begin{figure}
\includegraphics[height=30mm]{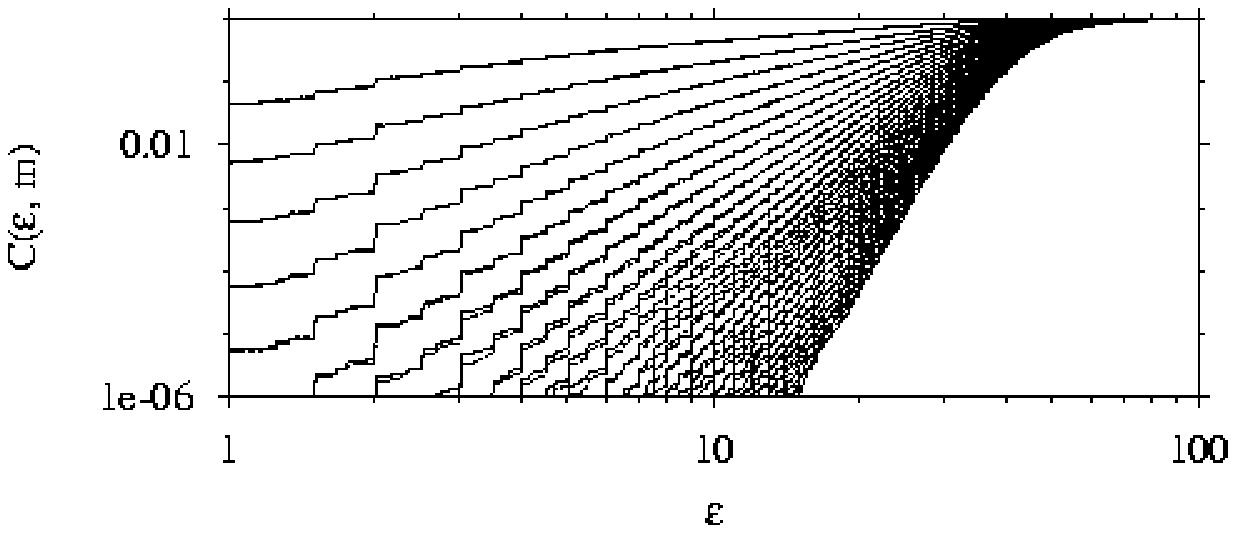}(a)
 \includegraphics[height=30mm]{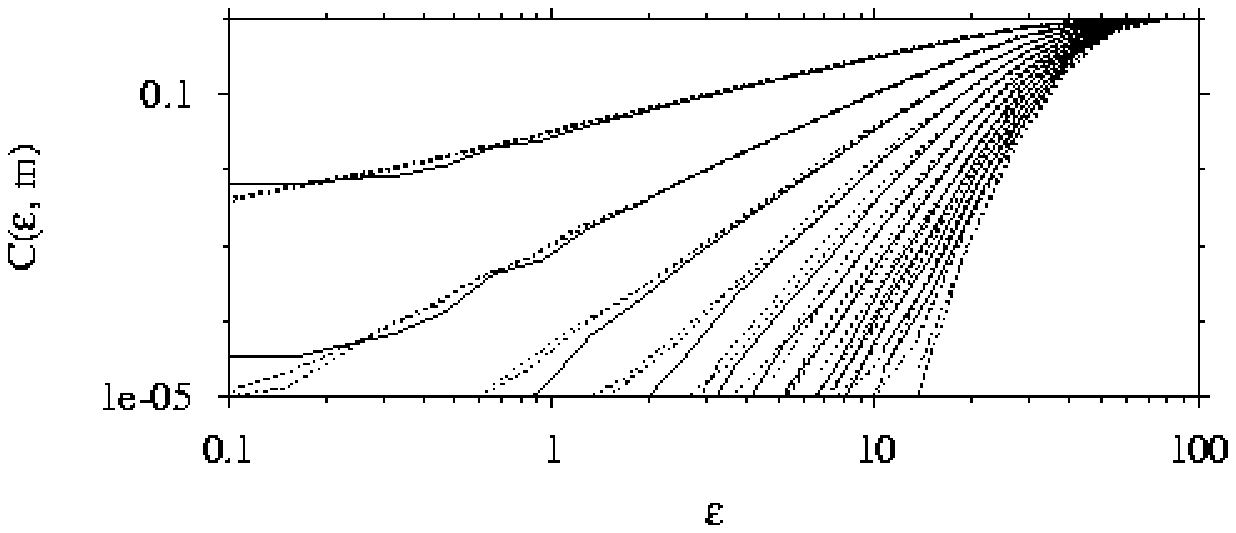}(b)
\caption{\label{siegel_fig19}
      \newline
         a) Grassberger-Procaccia plot for (a)
         freeflow- and (b) jammed
         state traffic data for varying dimensions
         and lag $\tau = 7$ seconds (solid lines).
         Grassberger-Procaccia with identical
         parameters for appropriate surrogate
         series realizations are plotted in dashed lines.}
\end{figure}
 Figure~\ref{siegel_fig19} shows Grassberger-Procaccia
 plots for (a) free flow traffic and (b)
 jammed state records for error distances
 $\epsilon$  of values $0.1\ldots 100$
 and embedding dimension $2 \ldots 25$.
 Both graphs fail to reveal any scaling region,
 moreover there is obviously no difference
 to Grassberger-Procaccia plots of
 appropriate surrogate realizations.\\
 The dimensions of merely stochastic systems
 appear infinite, therefore for this case it
 is a typical result, that the correlation
 integral reveals the embedding dimension (\cite{hegger}).

\subsubsection{Casdagli test}
 The local linear prediction of a time series
 in delay representation $\underline{x_t}$ is
 achieved by determination of a matrix $A$
 that minimizes the prediction error:
\begin{equation}
\sigma^2=\sum_{\underline{x}_t\in {\cal U}_t}
(\underline{x}_{t+1}-A\underline{x}_t -b_t)^2\label{eqno}.
\end{equation}
 where ${\cal U}_t$ denotes the $\epsilon-$
 neighbourhood of $\underline{x_t}$ excluding
 $\underline{x_t}$ itself.
 In some analogy to linear regression the
 prediction is:
 \begin{equation}\underline{x}_{t+1}^*= A\underline{x}_t+b_t\label{eqnp}.
 \end{equation}
 Local linear models are suggested
 a test for nonlinearity (\cite{casdagli}).
 The average forecast error is computed as
 function of the neighbourhood size on which
 the fit is performed. If the optimum occurs
 at large neighbourhood sizes, the data are
 (in this embedding) best described by a
 linear stochastic process, whereas an
 optimum at rather small neighbourhood sizes
 supports the idea of existence of a nonlinear,
 almost deterministic, equation of motion (\cite{hegger}).

\begin{figure}
\includegraphics[height=30mm]{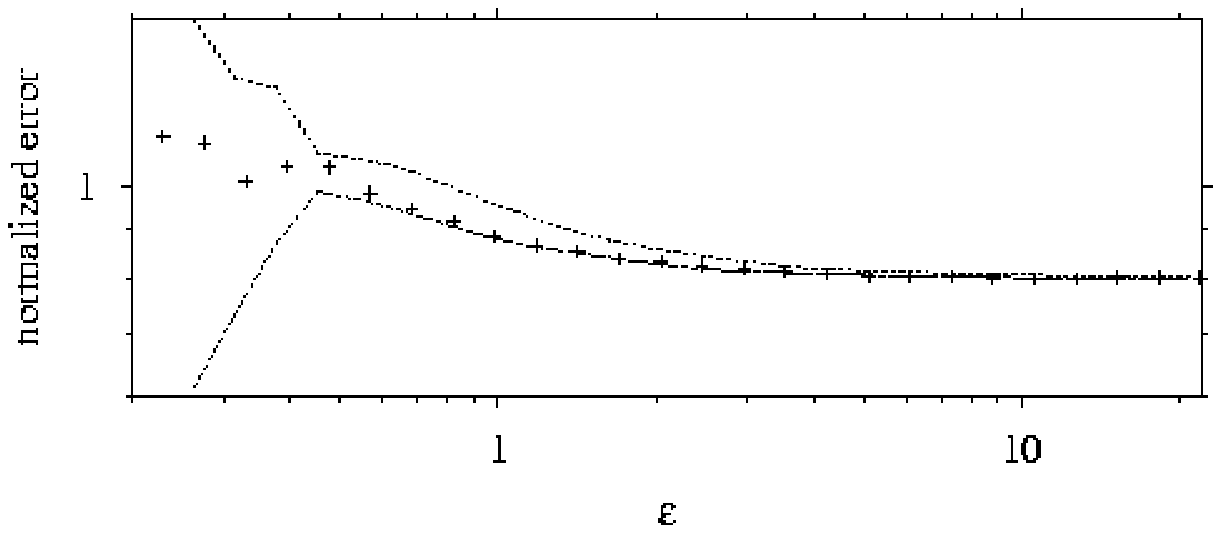}(a)
\includegraphics[height=30mm]{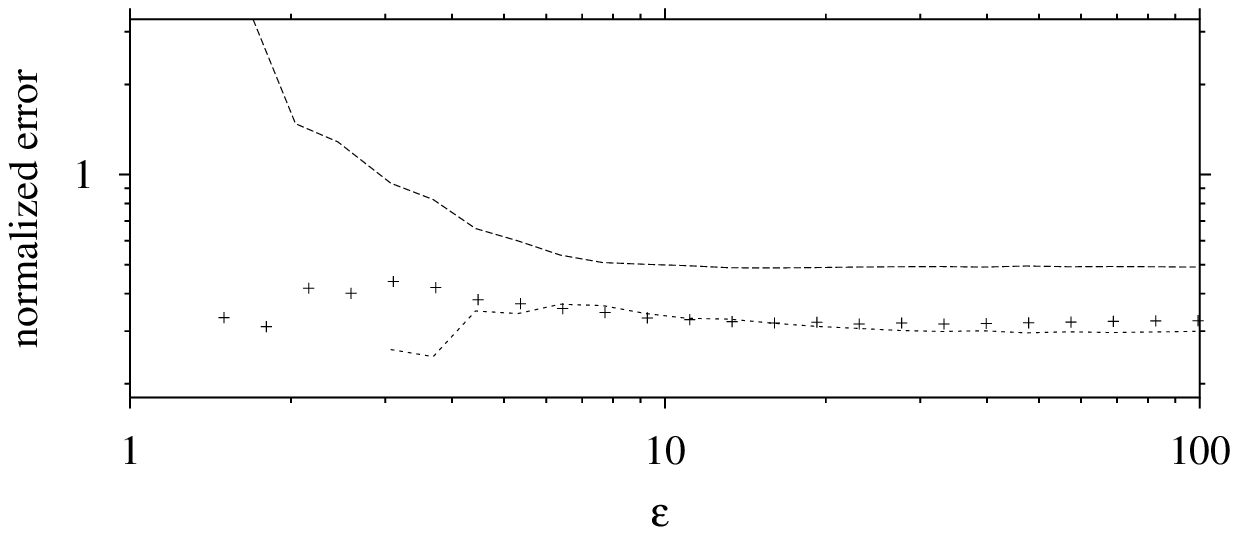}(b)
\caption{\label{siegel_fig21}
  Casdagli plot of (a) free-flow and (b) jammed state
  traffic data (crosses), compared to confidence bands,
  generated from to PRS series (dotted lines).}
\end{figure}
 In Fig.~\ref{siegel_fig21} the Casdagli-test was
 performed for original as well as for surrogate
 data. In the context of the Casdagli-test it is
 meaningless some of jam records fall not within 
 the surrogate-based confidence intervals.
 The qualitative comparability gives an
 indication for predominant stochasticity
 particularly in freeflow traffic records.

\section{Discussion}
 According to our analysis of traffic records,
 traffic dynamics is a two fixed point stochastic
 process, while the fixed points reflect
 the jam and freeflow regime.\\
 The abrupt transitions between the traffic
 states imply nonlinearity in the overall
 traffic dynamics. A variety of methods,
 more or less sensitive towards nonstationarity,
 yields Hurst-exponents that indicate
 long-range dependent dynamics in
 particular for freeflow traffic.
 Differenced as well as logdifferenced 
 velocity records reveal heavy tailed distribution, 
 however for both there is no clear scaling 
 region observabele to estimate scaling exponents.\\
 From our results it must be concluded that
 below diurnal time scales traffic data
 in jammed as in freeflow state exhibit
 neither deterministic nor low dimensional
 chaotic properties. \\
 The main intention of this article is to
 outline an overview of the stochastical
 properties achieved by data analysis
 of single-car road traffic records.\\
 Attending the problem of criticality in
 road traffic records, we find that as 
 well  the two fixed-point dynamics as the
 distribution of (differenced) velocities
 are contrary to the typical features of
 processes governed by self-organized criticality.
 This lets us rather suspect the rise of
 jams in the context of a (eventually, 
 but not necessarily, critical) phenomenon
 linked to a phase-transition.
 For such a model hypothesis, known  e.g.
 in equilibrium thermodynamics, the point
 of transition can be reached by fine
 tuning of a parameter.
 This must be distinguished from
 self-organized criticality, which represents
 the classification for systems attracted
 permanently by variable critical states.\\
 Contrary to the well-known conceptual analogy
 between traffic and granular flow, we rather
 propose an intuitive analogy of traffic dynamics
 with the condensation of steam to water.
 In contrast to the condensation of water
 driven by withdrawal of heat, free flow
 traffic condenses to higher particle density
 by an increase of trafficants in this picture.
 This increase can be interpreted as rised pressure.
 In accordance to such considerations
 and the empirical results of \cite{helbing}
 increasing traffic load (or input to the motorway)
 produces a (in the more popular sense) ''critical'' 
 tension that relaxes  in an abrupt transition to a jam.
 In this instructive example, traffic accidents,
 construction sites or slow vehicles
 could act comparable to condensation cores
 by exerting strong nonlinear negative
 feedback on the upstream traffic.
 The fine tuning parameter thus is
 the capacitiy of the motorway, limited
 by traffic load,  accidents or  construction sites.

\section*{Acknowledgments}
 The authors wish to thank Sergio Albeverio,
 Nico Stollenwerk and Michael Schreckenberg
 for fruitful discussions and the
 Landschaftsverband Rheinland
 (Cologne) for providing the data. This project
 was supported by the Deutsche Forschungsgemeinschaft
 (DFG), Sonderforschungsbereich 1114.
 We acknowledge the benefits of the TISEAN
 software package (available from www.mpipks-dresden.de).

\bibliography{siegelbonn2}

\end{document}